# PRECISION POINTING OF ANTENNAS IN SPACE USING ARRAYS OF SHAPE MEMORY ALLOY BASED LINEAR ACTUATORS

Nikhil S. Sonawane[*] and Jekan Thangavelautham[†]

Space systems such as communication satellites, earth observation satellites and space telescopes require precise pointing to observe fixed targets over prolonged time. These systems typically use reaction-wheels to slew the spacecraft and gimballing systems containing motors to achieve precise pointing. Motor based actuators have limited life as they contain moving parts that require lubrication in space. Alternate methods have utilized piezoelectric actuators. This paper presents Shape memory alloys (SMA) actuators for control of a deployable antenna placed on a satellite. The SMAs are operated as a series of distributed linear actuators. These distributed linear actuators are not prone to single point failures and although each individual actuator is imprecise due to hysteresis and temperature variation. The system as a whole achieves reliable results. The SMAs can be programmed to perform a series of periodic motion and operate as a mechanical guidance system that is not prone to damage from radiation or space weather. Efforts are focused on developing a system that can achieve one degree pointing accuracy at first, with an ultimate goal of achieving a few arc seconds accuracy. Bench top models of the actuator system has been developed and working towards testing the system under vacuum. A demonstration flight of the technology is planned aboard a CubeSat.

## INTRODUCTION

Satellites are playing an ever critical role, enabling inter-continental communication, monitoring human and natural threats, providing global-positioning services, observing Earth's changing weather and monitoring human impact on the environment. Exploratory probes and space telescopes[17,18] have been providing detailed observations of planets, moons and small-bodies throughout the solar-system, in addition to aiding in the discovery of stars, galaxies and exoplanets[18].

These systems are typically large and require use of actuators to point equipment including imagers, spectrometers, and communication antennas with a high degree of accuracy, at arcseconds or less. The actuators are typically large, expensive and depend on motors, gears and position control electronics. Due to the microgravity conditions of space, the actuators need to be especially lubricated using a unique class of dry lubricants[14]. These factors typically limit the life of the actuators, particularly metal to metal contact (tribological interactions) and make them vul-nerable to degradation through extended use, radiation and space weather events. While advancements are being made, tribology remains a major research topic in space technology.

---

[*] Master's Student, Space and Terrestrial Robotic Exploration Laboratory, Arizona State University, 781 E. Terrace Mall, Tempe, AZ.
[†] Assistant Professor, Space and Terrestrial Robotic Exploration Laboratory, Arizona State University, 781 E. Terrace Mall, Tempe, AZ

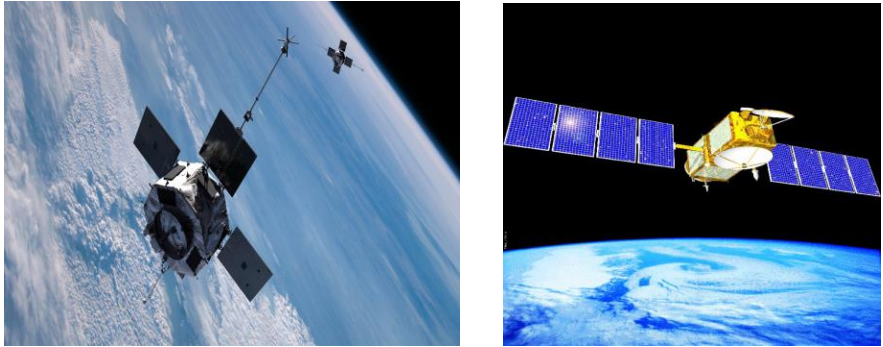

**Figure 1. Space probes and satellites may utilize solar panels, imagers, telescopes and communication antennas that require precise pointing.**

In this paper, we present Shape Memory Alloys (SMAs) as a replacement to conventional motors, servos and linear actuators for use on spacecrafts and space robots[15]. SMAs are typically metal alloys that transition from one phase to another due to relatively small changes in temperature. The alloy changes phase and thus can take on one or more structures. We exploit these natural physical phenomena to design linear actuators that expand and contract using heat. The advantage of SMAs is that the electronic required to control these systems need to only regulate temperature, and as a result, they are simple. Simple electronics can be well designed to better handle space-weather events and high radiation. These actuators also permit entirely new decentralized architectures, where one or more redundant actuators interact to perform precise positioning control[16]. Even if one or more actuators fail, the system works albeit with degraded performance. This is in contrast to conventional actuators such as motors and servos that may only have one or two levels of redundancy and with individual component failure, the result is catastrophic failure[16].

Conventional linear actuators have been considered and they use motors and worm-gears that require dry lubricants[14]. These systems are complex and face many of the same challenges as conventional motors and servos. An alternative is electroactive polymers also known as Dielectric elastomer actuators (DEAs)[13]. DEAs have been used as artificial muscle; however, in space they are prone to rapid degradation due to UV exposure. Hence the polymer membrane needs to be well shielded. Another class of actuators includes piezoelectric actuators that expand or contract due to an electrical current. Piezoelectric actuators have been proposed to actuate Asteria, an exoplanet observing CubeSat[12]. However piezoelectric actuators have limited expansion capacity and hence are best for high-precision positioning.

We have found that having reviewed many credible actuator technologies, Shape Memory Alloys are relatively simple and robust but are widely misunderstood. However, the challenge with SMAs is understanding their behavior. One particular challenge faced is the inherent stochasticity of shape memory alloys that results in 5-10 % variability in performance. Our work focuses on the design of an effective mechanism design and accompanying controls approach to overcome the challenge of inherent stochasticity.

In the following sections, we present related work on actuators and precision pointing for space, followed by presentation of SMAs. We then present experimental design to precisely point a communication antenna using shape memory alloys. We analyze the results of the experiments, followed by discussions, conclusion and future work.

**BACKGROUND**

Conventional space systems utilize gimbals to enable precise pointing of instruments and antennas. Two sets of orthogonal gimbals are used for the rotation against azimuth and elevation axes. One example is SSTL Antenna Pointing Mechanism (APM)[4] ,which has a step size ≤ 0.024 deg. The gimbal based APMs work using a stepper motor giving very precise and fine steps for rotation of a pointing device. There are, however, several drawbacks to these systems including slow response time, limitation on the motion range and also singularities in positioning due to system design. Thermal vacuum testing has shown that high-accuracy is difficult to achieve. Nonlinearities, build-errors, non-orthogonality effects and thermal expansion effects combine to reduce the accuracy of conventional gimbal systems.[2]

Therefore, there is a need for an alternate, low-cost actuator to compete against the existing APM technology. In this work, we propose use of Shape Memory Alloys as linear actuators. SMAs have the tendency to return to its original memorized form when heated. In the 1990s, they were popularized as "smart metal" and used to make bendable eyeglass frames. SMA performs this "recall" due to phase transformation of the alloy molecular structure. Furthermore, corrective action can be done by "reprogramming" the shape memory alloy under high heat.[5]

SMAs are especially suited for small satellites and CubeSats when compared to other actuators because of the stringent mass and volume constraints. In addition, SMAs have some of the highest operating efficiency (work output) when compared to all the commercially used micro actuators[6]. SMAs have been designed for terrestrial use in linear-actuators, large force actuators, glass frames, couplings, rivets, clamps, vibration dampers, vascular stents used in heart surgery and as space qualified latching mechanisms[7,8,9]. However, there is yet to be space-qualified SMA for linear actuation and applications in precise pointing. Certain types of SMA that we present here in this paper have the technical potential to meet precise pointing requirements. However, these actuators are not persistent and require input heat or power to maintain precise positioning. Our work introduces a new mechanism that addresses the inherent limitations of SMAs to achieve precise pointing.

**ACTUATOR DESIGN OVERVIEW**

In this section, we present efforts to develop linear actuators using shape memory alloys. Factors to consider in designing the linear-actuator are high load bearing capability, low-cost, simple construction, reliable actuation and enabling one degree overall point accuracy or less. First, we analyze the deflection and load bearing behavior of SMAs when heated (Figure 2).

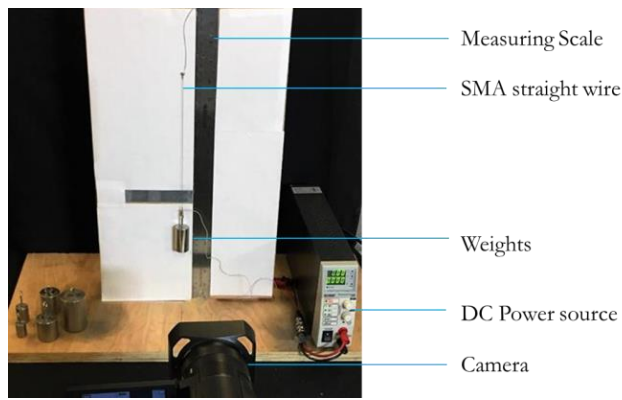

**Figure 2. Experimental setup for SMA wire characterization.**

These experiment results formed the basis of the initial design parameters to be considered for designing the mechanism.

**Mechanism Selection**

SMA wires can handle large stresses, but give very small deflections. These SMA wires were found to only repeatably contract/expand 3-4%. This was consistent with previously obtained laboratory results[10]. To increase the effective stroke length, a simple lever mechanism[11] was chosen. The lever translates the high mechanical pull force into a large deflection. Furthermore, use of a lever allows flexibility in choosing stroke length, without modifying the SMA wire. After the lever has achieved a desired expansion, a latching mechanism is used to "lock" the expansion. Otherwise, heat will need to be constantly supplied to maintain the extended SMA wire length. The latching mechanism used is a Right Angle Pull Mechanism as shown in Figure 3. This mechanism was selected based on its reliability and wide use[11].

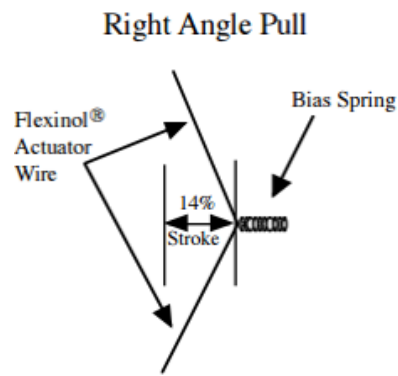

**Figure 3. Right angle pull mechanism.**

**System Design Parameters**

The major parameters considered while designing our actuation system is described below[10]:

*SMA Diameter* - The SMA selected was Nitinol®, with a wire diameter of 0.015 in for the linear actuator 0.002 in for the latching mechanism. The selected diameter determines the maximum break force, the cooling rate and minimum bending radius.

*Temperature* – The transition temperature, the temperature at which the wire starts contracting is 90°C

*Initial Shape Programming* – The initial shape programmed through heating determines the amount of strain a wire can withstand. For our experiments, we use a straight wire 20 cm in length that can withstand 3-4% contraction/expansion and sustain a million cycles.

*Stroke Length* – Each linear actuator needs to have a high enough pulling force and stroke length sufficient to achieve maximum tilt angle.

*Environmental Conditions* - The other materials in contact with the SMA wire act as a heat sink. The rate at which it is cool determines the overall cycle rate. In space, heat dissipation needs to be carefully considered, as convection cooling doesn't exist. Instead, heat needs to be dissipated via conduction using a cold sink.

*Stress/Strain Behavior* - The maximum strain imposed on the SMA is lowered to maximize reliability and repeatability. For 8% strain, SMA only last a few cycles, while for 5-8%, it will work for a thousand cycles and at 3-4% strain, it will operate for a million cycles.

*Bias Force* - The bias force is the force required to deform or extend a cooled SMA wire to bring it back into a 'detwinned martensite' (reset) state. For the chosen wire it is 8.1 N.

*Cycle Rate* – The cycle rate is the time required to heat the SMA wire, extend the linear actuator and lock in the extension. In our experiments, it takes 3 secs.

**PROTOTYPE DESIGN**

Figure 4 shows a CAD model of the Actuator Pointing Mechanism (APM) utilizing three SMA-based linear actuators. Many linear actuators maybe added to increase redundancy and pointing precision. This prototype (excluding the parabolic antenna) is 43 cm in height and has a width of 36.5 cm.

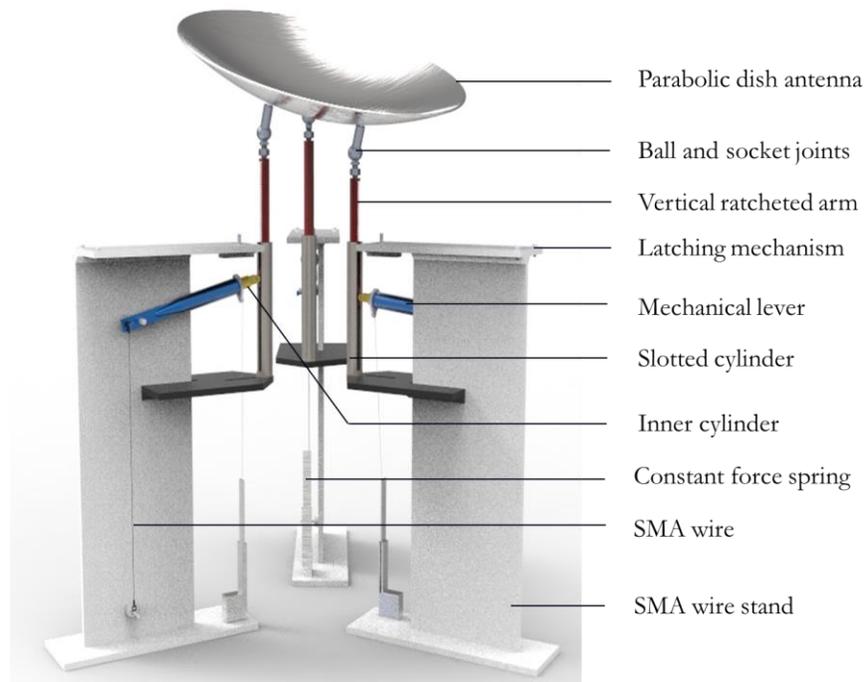

**Figure 4. CAD model of the proposed actuator pointing mechanism using 3 linear actuators.**

For each linear actuator, the shorter end of the lever is attached to one end of the SMA wire, while the other end is fixed to the body. The SMAs have a natural tendency to contract on heating, which tends to pull the lever downwards and rotate about the fulcrum. As can be seen in the Figure 4, to convert this circular motion into linear motion, the longer end of the lever has a telescopic cylinder that only captures linear motion. The inner cylinder of the telescopic body has a pin joint with a vertical ratcheted arm having the ability to oscillate inside a slotted cylinder. The vertical ratcheted cylindrical column on the other end is attached to the parabolic dish antenna as shown.

Each linear-actuator contributes to tilt and yaw of the antenna. More actuators maybe added to increase redundancy and precision. Control of the antenna pointing angle is achieved by adjusting three control points. The elevation of these points is controlled using SMA actuated movement. The maximum angle of tilt was set at 45° according to which the stroke ratio required was 1:10 (Figure 5).

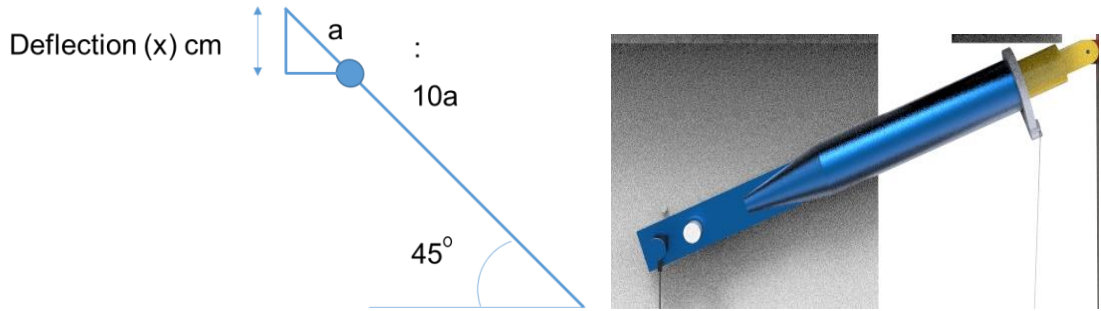

**Figure 5. Schematic and CAD model of the mechanical lever.**

Assuming the short and long lever arms to be of lengths *a* and *10a,* respectively, from Figure 5, applying basic trigonometry:

$$\Delta d(x) = a \sin 45^o \qquad (1)$$

Using an 8.1 N bias force, we get 0.7 cm deflection, which is 3.5 % of the total wire length of 20 cm. Furthermore, by similarity of triangles the deflection is magnified 10 times due to the lever ratio. With a maximum SMA deflection of 0.7 cm, the max angle that can be achieved is 41°. To attain 45° angular tilt for the antenna, the vertical actuators would have to be adjusted and placed further towards the periphery of the antenna. Figure 6 shows the system layout. Using trigonometry, the value of *y* is 7 cm. This is the distance between the virtual circles on the periphery of which, these actuator assemblies will be placed.

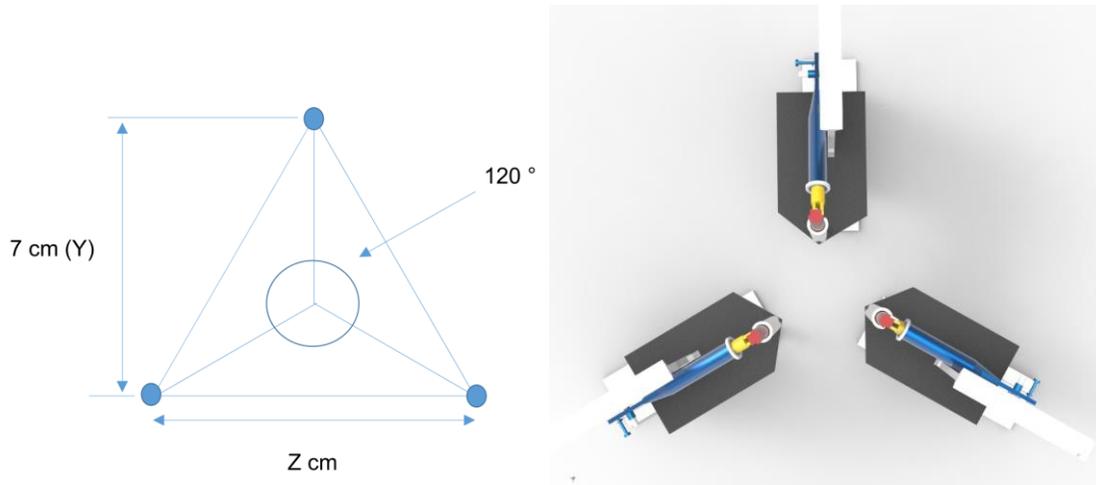

**Figure 6. Schematic and CAD of the 3 vertical arms.**

By trigonometry, z=8.1 cm as shown in Figure 6. Therefore, the system defined will have the ability to move ±41° elevation with 360° azimuth angle by the simultaneous control of the 3 vertical actuators.

**Latching Mechanism**

The latching mechanism shown in Figure 7 locks the extending or contracting linear actuator without having to provide a constant source of heat (and power). This is achieved by retracting a latch pin. The vertical ratcheted arm has ratchets on its body based on the geometry observed from Figure 4. The 7 cm vertical distance is divided by 45 degrees to have ratchets at a distance 0.16 cm which is equivalent to 1 degree. Thus, antenna could be tilted at 1 degree increments.

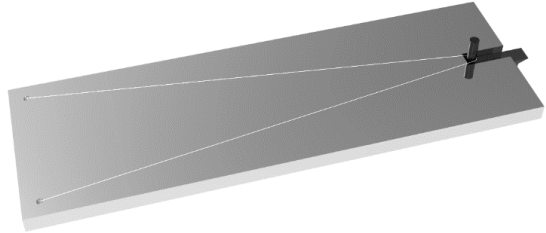

**Figure 7. Latching mechanism developed for the antenna pointing mechanism.**

## RESULTS AND DISCUSSIONS

The experiments performed on the SMA wire as shown from Figure 2 were aimed to obtain the maximum deflection value at bias force for the chosen wire i.e. 8.1 N. The value corresponding to a 8.1 N bias force is a 0.7 cm extension of the SMA wire. Figure 8 shows a non-linear increase in pull force due to extension of the SMA wire. Too high a deflection can damage the actuator, as seen from this graph. As a result, a precise controller is required for heating the SMAs to attain the desired extension.

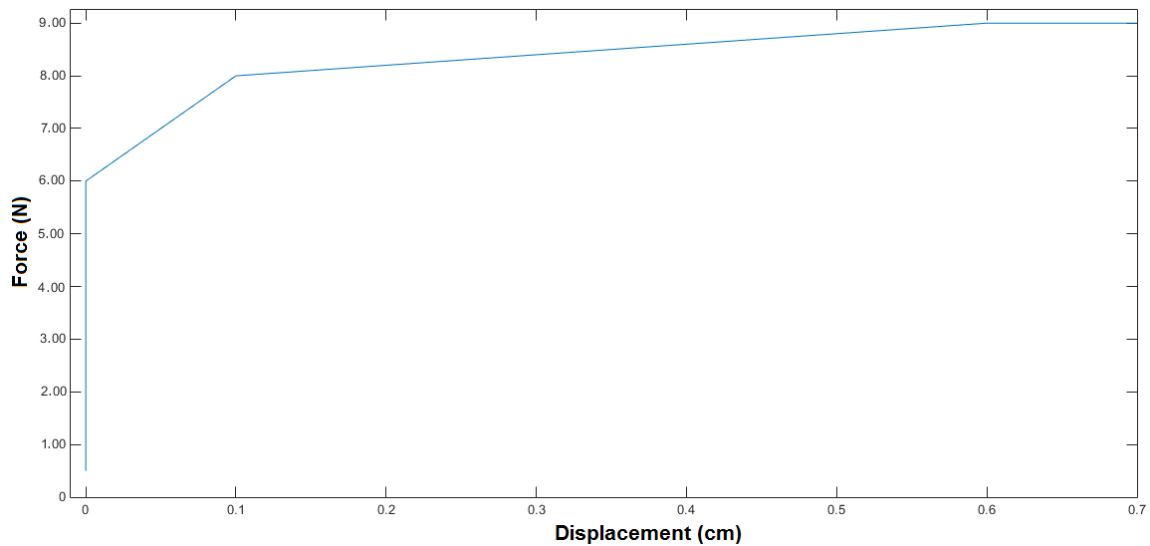

**Figure 8. Force vs. displacement for a Nitrol® SMA Wire.**

**Demonstration of SMA based pointing and latching mechanism**

The laboratory prototype uses a DC power source to heat the SMA wire and thus trigger the actuation. As timing is critical to achieve an accurate response, an Arduino microcontroller was used to provide an accurate timed signal to trigger the linear actuator, followed by triggering the latch actuator. Figure 9 shows system diagram for the experimental system and Figure 10 shows the major components used in our experiments. Commanding is done using a microprocessor that triggers a relay which switches on a heater, thus activating a linear actuator, followed by timed triggering of the latch to lock in the linear actuator position.

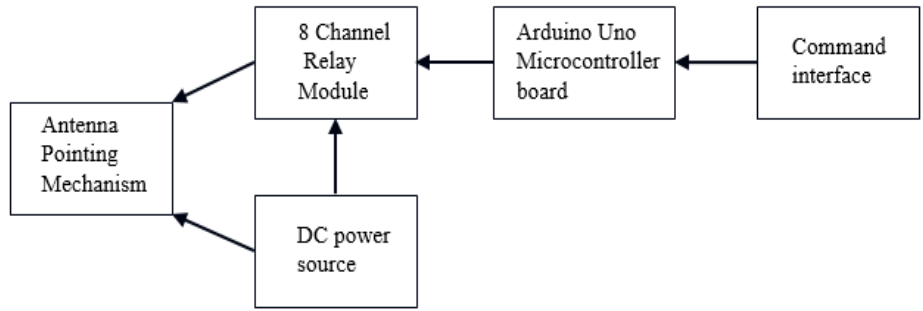

**Figure 9. System interface for experimental setup.**

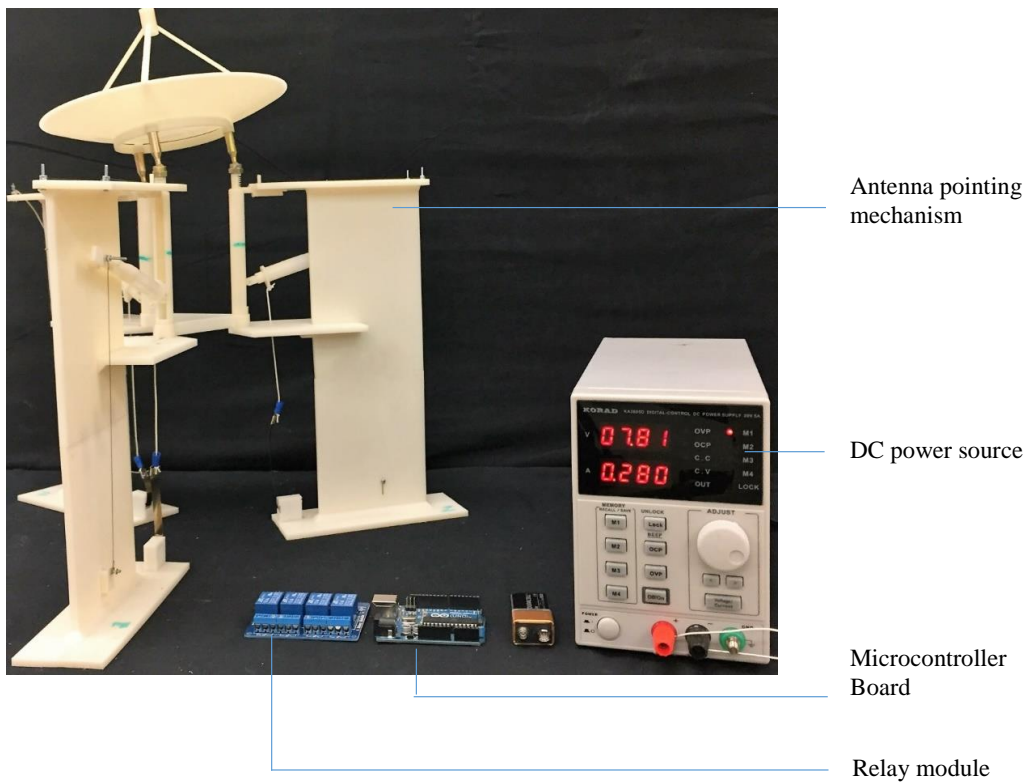

**Figure 10. Experimental setup components**

**Working Mechanism for 1 Cycle**

Figure 11 shows the full sequence of steps to reorient the communication antenna, starting with (a) Latching, (b) Unlatching, (c) Initial zeroing of antenna, (d) Setting of stroke (with overshoot), (e) Downward motion of vertical arm to lock linear actuator, (f) Latching of actuator at desired angle.

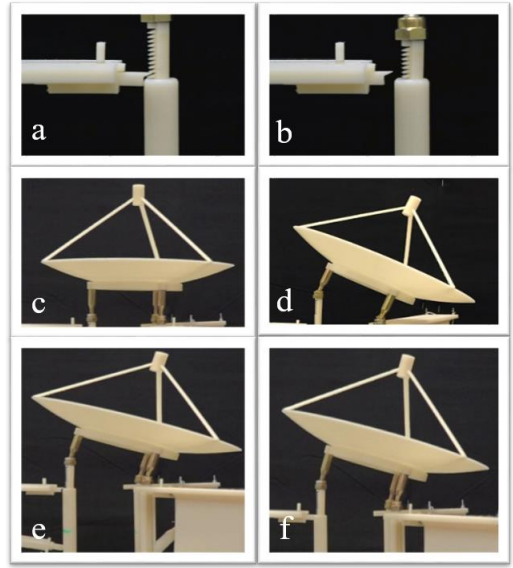

**Figure 11. Full working cycle of SMA antenna pointing mechanism, from (a) through (f).**

**System and Subsystem Repeatability**

In this section we analyze the repeatability of our actuator pointing mechanism over several cycles (Figure 12). For the experimental setup, the required angle depends on the time at which the latching takes places along with the movement of the linear actuator. Experiments were performed to analyze the repeatability and change in behavior of the latching mechanism for varying latch time delays with current being supplied to the wire for 1 second.

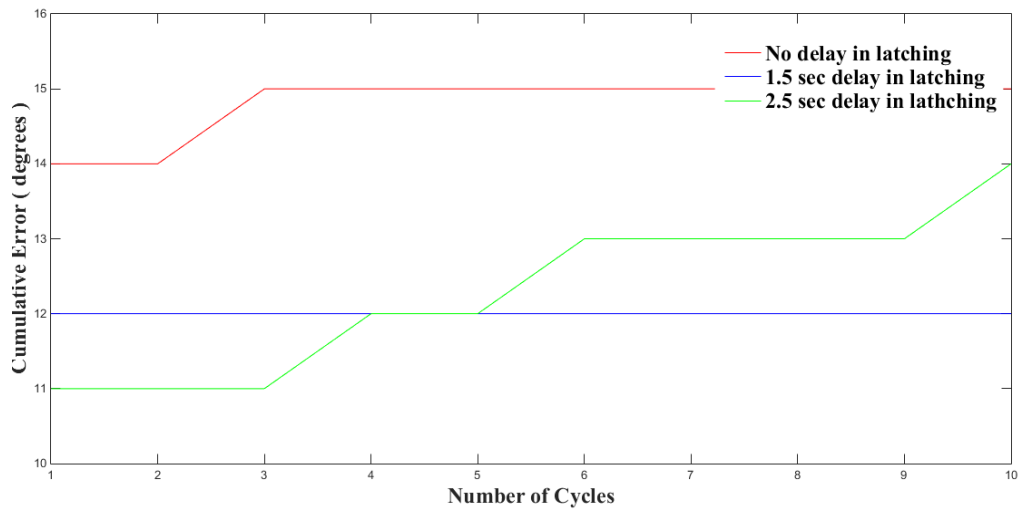

**Figure 12. Cumulative error for increasing number of system actuation cycles.**

As shown in Figure 12, with no delay in latching, there is a 1-degree cumulative error over 10 cycles. Similarly, in the case of a 1.5 second delay in latching, no error is observed over a period of 10 cycles. In the third case, for a 2.5 second delay in latching, the error propagates by three degrees cumulative for 10 cycles. Further, the uncertainty in the error of all the three cases is due to the error sources present in the system.

**Performance Analysis**

The linear movement of the 3 actuators was studied using antenna mockup. Angles achieved were compared to determine the performance of each actuator. The amount of time for which current was provided was varied in each case with errors being mainly due to placement of actuators and lack of manufacturing/design precision. Figure 13 shows how error is propagated with increased heating time of the SMA actuators. It is critical to heat the SMA to the target temperature, but do so as quickly as possible.

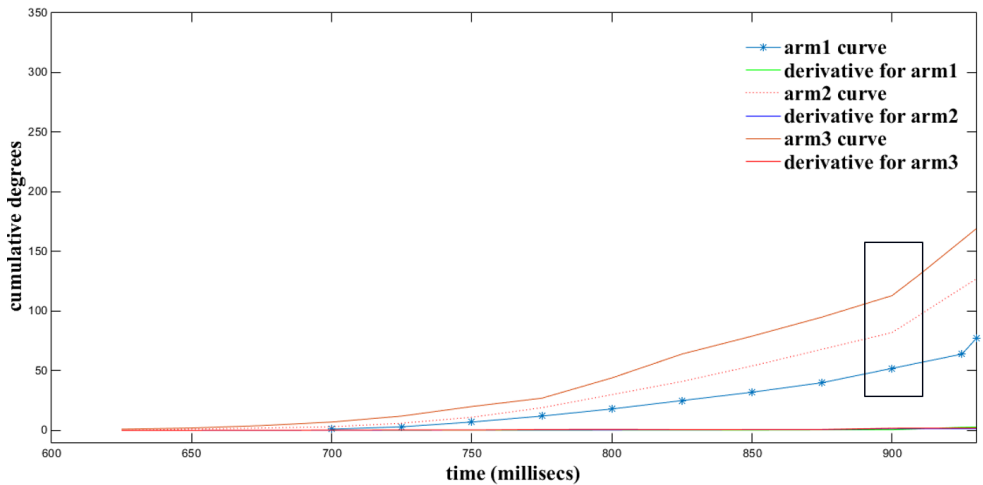

**Figure 13. Relative variation in error propagation due to settling time.**

To study the rate at which the angle changes, we determined the 1st order derivative of all the curves after using basic curve fitting tool in MATLAB. The results in all the 3 cases show increasing slope line, signifying increasing change in the angle achieved. The sudden change in the slope of the curve represents the transition temperature range for the SMA wire. It works on the principle of Joule's effect and, thus, as the time approaches 1 sec, the transition temperature range is reached and sudden contraction in the SMA wire is achieved.

The above discussed results are based on experiments performed to test the proper working of the proposed APM. The errors in the results obtained are attributed to factors like design and manufacturing errors, heating/cooling of SMA wire, positioning/working of the constant force spring and placement of actuators. The tilt angle achieved depends on the contraction of the SMA wire, the activation temperature of the wire, the amount of time for which it is heated and time for which it is cooled. The amount of time for which the current is provided is precisely controlled by a micro-controller and this helps achieve excellent repeatability. The manufacturing errors are attributed to extrusion molded 3D printing, which limits the surface smoothness by 0.1 mm. Precise manufacturing techniques with Ceramic type materials for construction will help reduce errors.

Overall, SMAs shows promise as linear actuators. Furthermore, these linear actuators maybe applied for precise pointing of antennas and other sensors. The experiment system is limited to one degree accuracy, but this can be improved through precise manufacturing.

The SMAs prove to be ultra-low-mass and have large load carrying capacity. They have large transitional (operating) temperature range, which makes them suitable for use in space. Further, work is required to understand how to limit error propagation and speed up actuation.

**CONCLUSION**

In this work, we propose use of shape memory alloys for use in linear actuators for precision pointing applications. Multiple linear actuators are combined to control the pointing angle of an antenna. Shape memory alloys can have high extension/contract, but at the price of poor reliability and repeatability. Our work shows that with sufficiently small extension/contraction, the results are expected to repeatable for a million cycles and have excellent reliability. To compensate for the small extension/contraction, we utilize a lever mechanism to effectively increase the stroke length. In addition, our design uses a latching mechanism to set the stroke length, thus simplifying actuator control and repeatability. Using this approach, $1^{o}$ pointing accuracy was achieved for a laboratory prototype. Precision machining can be used to further increase the point accuracy. This work shows a promising pathway towards development of a space-qualified antenna pointing system using shape memory alloys.

**REFERENCES**

[1] R.L. McNutt, "Space Exploration." *Handbook of Space Engineering, Archaeology, and Heritage*. CRC Press, 2009. 835-855.

[2] A.J.D. Brunnen, and R. H. Bentall. "Development of a high stability pointing mechanism for wide application." (1982).

[3] "NASAFacts." h*ttps://www.nasa.gov/centers/langley/pdf/70810main_FS-1996-08-09-LaRC.pdf*. Langley Research Center Hampton, Virginia 23681 Office of Public Affairs, Web.

[4] M.Ferris and N.Phillips, "The use and advancement of an affordable adaptable antenna pointing mechanism." *Proc. 14th Eur. Space Mech. Tribol. Symp*. 2011.

[5] W.Huang, "Shape memory alloys and their application to actuators for deployable structures." (1998).

[6] F.Butera, A.Coda, G.Vergani, and S.G. SpA, "Shape memory actuators for automotive applications." *Nanotec IT newsletter. Roma: AIRI/nanotec IT* (2007): 12-6.

[7] J.R.S. Anadon, "Large force shape memory alloy linear actuator." PhD diss., University of Florida, 2002.

[8] D.J. Hartl and D. C. Lagoudas. "Aerospace applications of shape memory alloys." *Proceedings of the Institution of Mechanical Engineers, Part G: Journal of Aerospace Engineering* 221.4 (2007): 535-552.

[9] Aerospace, TiNi. "Frangibolt." (2015).

[10] Dynalloy, I. "Technical characteristics of flexinol actuator wires." *CA: Tustin* (2011).

[11] R.G. Gilbertson, *Muscle Wires: Project Book;* Mondo-tronics, 2000.

[12] M. W. Smith, S. Seager, C. M. Pong, M. W. Knutson, D. W. Miller et al., "The ExoplanetSat Mission to Detect Transiting Exoplanets with a CubeSat Space Telescope," *25th Annual AIAA/USU Conference on Small Satellites*, Logan, Utah, 2016, pp. 1-9.

[13] J. S. Plante, S. Dubowsky, "Large-scale failure modes of dielectric elastomer actuators," International Journal of Solids and Structures, Vol. 43, No. 25–26, 2006, pp. 7727–7751.

[14] R.L. Fusaro, "Lubrication of Space Systems," NASA Technical Report, 106392, 1994.

[15] Thangavelautham, J., Abu El Samid, N., Grouchy, P., Earon E., Fu, T., Nagrani, N., D'Eleuterio, G.M.T., "Evolving Multirobot Excavation Controllers and Choice of Platforms Using Artificial Neural Tissue Controllers," Proceedings of the IEEE Symposium on Computational Intelligence for Robotics and Automation, 2009.

[16] A. M. Bilton, S. Dubowsky, "Inverse Kinematics for the Control of Hyper-Redundant Binary Mechanisms with Application to Solar Concentrator Mirrors," Latest Advances in Robot Kinematics, pp. 421-428, 2012.

[17] G. A. Beals, R. C. Crum, H. J. Dougherty, D. K. Hegel, and J. L. Kelley. "Hubble Space Telescope precision pointing control system", *Journal of Guidance, Control, and Dynamics*, Vol. 11, No. 2 (1988), pp. 119-123.

[18] N. M. Batalha, "Exploring exoplanet populations with NASA's Kepler Mission," *Proceedings of the National Academy of Science*, vol. 111, no.35, pp. 12647–12654, 2014.